\documentclass[runningheads]{llncs}

\usepackage{cite}
\usepackage{amsmath,amssymb,amsfonts}
\usepackage{algorithmic}
\usepackage{graphicx}
\usepackage{textcomp}
\usepackage{xcolor}
\usepackage{adjustbox}
\usepackage{balance}
\usepackage{enumitem}
\usepackage{subcaption}
\usepackage[T1]{fontenc}
\usepackage[utf8]{inputenc}
\usepackage{booktabs}
\usepackage{orcidlink}

\usepackage{xspace}
\newcommand{\proposedapp}{ReBoT-MiCS\xspace}

\usepackage{fancyhdr}

\fancypagestyle{firstpage}{
  \fancyhf{} 
  \fancyfoot[L]{\textcolor{blue}{%
    \parbox{0.99\textwidth}{\footnotesize
    This paper is accepted to be published in the 34th International Conference on Real-Time Networks and Systems (RTNS) 2026.
    }%
  }}

}

\def\BibTeX{{\rm B\kern-.05em{\sc i\kern-.025em b}\kern-.08em
    T\kern-.1667em\lower.7ex\hbox{E}\kern-.125emX}}

\begin{document}
\title{\proposedapp: \underline{Re}liability-Aware \underline{Bo}unding \underline{T}ime for Enhancing Timing Behavior of Embedded \underline{Mi}xed-\underline{C}riticality \underline{S}ystems\\
}

\titlerunning{ReBoT-MiCS: Reliability-Aware Bounding Time of MC Systems}

\author{Mohammad Abbasinia\orcidlink{0000-0001-6056-1465}\thanks{Corresponding Author}, Behnaz Ranjbar\orcidlink{0000-0001-7944-7101} 
\and Akash Kumar\orcidlink{0000-0001-7125-1737}}

\authorrunning{M. Abbasinia, B. Ranjbar and A. Kumar}

\institute{Chair of Embedded Systems, Ruhr University Bochum, Germany\\
\email{\{mohammad.abbasinia, behnaz.ranjbar, akash.kumar\}@rub.de}\\}

\maketitle

\thispagestyle{firstpage}

\begin{abstract}
In Mixed-Criticality (MC) systems, considering multiple Worst-Case Execution Times (WCETs) for the same task, corresponding to different system operation modes, presents a fundamental challenge in MC system design. Identifying suitable WCETs for lower (LO) criticality system operational modes is inherently difficult. 
Tight low WCET values may trigger frequent mode switches that degrade Quality-of-Service~(QoS) by dropping/degrading the lower-criticality tasks and reducing their functionality.  However, more conservative values reduce resource efficiency. Moreover, low WCET selection affects exposure assumptions under the adopted transient-fault model, which is used in reliability-aware scheduling analysis. 
Although some studies have considered low WCET determination to make a reasonable trade-off between processor utilization, mode switching probability and QoS, reliability constraints are typically addressed separately through fault-tolerance mechanisms. This separation overlooks the opportunity to use execution-time bounds as part of design-time reasoning about exposure-related reliability constraints. 
In this regard, we propose a reliability-aware analytical scheme, named \textit{\proposedapp}, to determine low WCETs at design time, jointly balancing utilization, QoS, mode switches, and probabilistic reliability constraints of each task criticality level under a transient-fault exposure model. Experiments conducted with various synthetic and embedded benchmarks demonstrate the efficacy of the proposed scheme, in which utilization can be improved by up to 27.6\% and 11.67\% on average, while bounding the mode switching probability by up to 5.48\%, and satisfying the specified reliability constraints.
\end{abstract}

\keywords{
Mixed-Criticality, Mode Switching Probability, Quality-of-Service, Reliability-Constrained Design, Timing Analysis, Utilization.
}

\section{Introduction}
\label{sec:Intro}
Mixed-Criticality (MC) systems have become prominent in modern safety-critical domains, including automotive and avionics, where tasks with different criticality levels are executed on a shared common platform~\cite{Baruah2012,Guo2018,Liu2016,Ranjbar2021Date,burns2026mixed}. In such MC systems, tasks are associated with multiple Worst-Case Execution Time~(WCET) estimates, corresponding to different system operational modes. In lower-criticality modes (LO modes), tasks are scheduled considering the WCET obtained for the LO mode~($\textit{WCET}^{\textit{LO}}$). If at least one High-Criticality (HC) task exceeds its ~$\textit{WCET}^{\textit{LO}}$, the system switches to a higher-criticality mode (HI mode), in which HC tasks are scheduled with their highest WCET values~($\textit{WCET}^{\textit{HI}}$). In this mode, Low-Criticality (LC) tasks are often treated with reduced service in order to preserve the HC timing guarantees~\cite{Baruah2012,Liu2016,Ranjbar2021Date}. In the classical MC system model, this effect may be represented by dropping LC tasks; however, in many realistic systems, a more acceptable design choice is to continue serving LC tasks with a degraded but still useful level of service.

While using multiple WCETs and mode switching is a key factor in MC system design, it introduces fundamental challenges. A lower $\textit{WCET}^{\textit{LO}}$ can improve processor utilization by allowing more LC tasks to be executed in LO mode. However, it increases the probability of mode switching, as HC tasks are more likely to exceed their $\textit{WCET}^{\textit{LO}}$ bounds and overrun. Conversely, considering a higher $\textit{WCET}^{\textit{LO}}$ for HC tasks reduces the probability of mode switching, thereby improving Quality-of-Service~(QoS) by less dropping/degrading LC tasks in HI mode, albeit at the expense of utilization~\cite{Ranjbar2021Date,Ranjbar2022BotMiCS}. Some studies on determining $\textit{WCET}^{\textit{LO}}$ have therefore focused~on balancing the trade-off between utilization and mode switching behavior. This trade-off highlights that $\textit{WCET}^{\textit{LO}}$ is not merely a timing parameter, but a critical design variable that directly affects resource efficiency, mode switching behavior, and the service delivered by LC tasks. 
A key aspect of this trade-off is the treatment of LC tasks after a mode switch. While dropping some/all LC tasks simplifies MC schedulability analysis, it can be overly pessimistic in practical systems where LC services may still provide useful outputs at reduced quality. 
This motivates degradation-based MC models, in which LC tasks are not necessarily dropped in HI mode but remain active with reduced budgets and degraded, yet acceptable, service levels~\cite{Guo2018, Liu2016, Liu2018}. Such degradation is applicable to services such as image processing, monitoring, perception, communication, and signal processing, where approximate or lower-quality outputs may still be useful~\cite{Liu2016, Liu2018, axbench}.

In addition to ensuring real-time constraints, satisfying reliability requirements is also vital in MC system design~\cite{burns2026mixed,Huang2014}. Safety standards often express such requirements using metrics such as Probability of Failure per Hour (PFH), which quantifies the rate at which failures may occur during system operation; thus, a lower PFH, i.e., higher $-\log_{10}(PFH)$, indicates better reliability~\cite{Navardi20222,koren2020fault}. Reliability in MC systems is usually addressed through hardware robustness or explicit fault-tolerance techniques that introduce redundancy into the system~\cite{burns2026mixed}. Although these techniques can improve fault coverage, they introduce timing, memory, energy, and schedulability overheads. In real-time systems, such overheads are particularly important because reserving additional time for recovery or redundant execution may reduce the utilization available to LC tasks, increase timing analysis pessimism, or even make the task set unschedulable. This motivates the question of whether exposure-related reliability constraints can be considered earlier in the MC design flow, before applying more expensive fault-tolerance mechanisms.

\begin{figure}[t]
\centerline{\includegraphics[width=0.9\columnwidth]{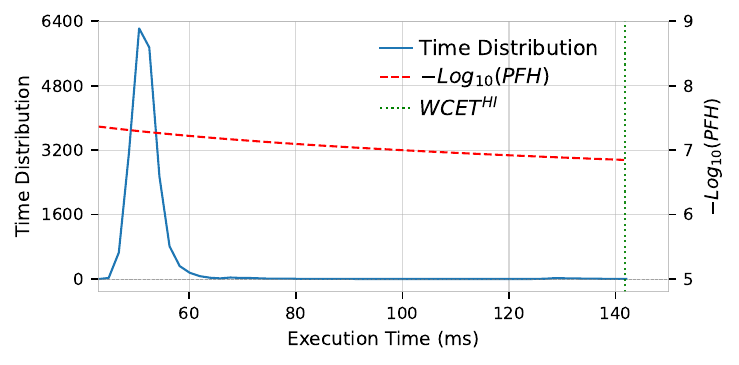}}
\caption{Measured execution-time distribution of the insertion-sort benchmark from MiBench~\cite{Mibench} on ODROID-XU4, with corresponding analytically computed PFH bounds under the adopted transient-fault model.}
\label{fig:run app on a board}
\end{figure}

This paper takes a reliability-aware view of $\textit{WCET}^{\textit{LO}}$ determination. The key observation is that, under a stated transient-fault model, the execution-time budget used in the analysis defines a conservative fault-exposure window for a task. Therefore, selected $\textit{WCET}^{\textit{LO}}$ values can be evaluated not only with respect to schedulability, utilization, and mode switching probability, but also with respect to whether it satisfies a PFH constraint. Importantly, this does not imply that changing $\textit{WCET}^{\textit{LO}}$ changes the physical fault rate or the actual execution-time distribution of a task. Instead, $\textit{WCET}^{\textit{LO}}$ is used as a design-time exposure budget for deriving conservative PFH bounds. 
Under this interpretation, larger $\textit{WCET}^{\textit{LO}}$ values do not physically make a task less reliable; rather, they 
increase the analyzed exposure window that must satisfy the PFH constraint, whereas smaller values may reduce the exposure bound but increase the probability of mode switching and QoS degradation.
This creates a design trade-off among utilization, mode-switch behavior, QoS, and PFH feasibility.

To make this trade-off concrete, consider an individual task in an embedded system, where the task execution times vary across job instances. Its measured executions can be used to characterize its timing, such as the execution-time distribution. 
Fig.~\ref{fig:run app on a board} illustrates this intuition using an embedded benchmark from ~\cite{Mibench} executed on an ODROID-XU4 board. 
The corresponding failure probabilities are then computed analytically from the assumed adapted transient-fault model and selected exposure window; they are not measured rare hardware-failure rates.
As illustrated, different candidate execution-time bounds for $\textit{WCET}^{\textit{LO}}$ lead to different conservative exposure-based PFH bounds. 
For example, if $\textit{WCET}^{\textit{LO}}$ is chosen from the interval of $(100,\textit{WCET}^{\textit{HI}}]$, 
then ${-\log_{10}(PFH)}|_{({100<\textit{WCET}^{\textit{LO}}\leq\textit{WCET}^{\textit{HI}}})}$$<$7. In the case where the PFH constraint being $10^{-7}$ ($-\log_{10}(PFH_{const})$=7), we cannot guarantee it, since the calculated PFH violates the PFH constraint. In contrast, if $\textit{WCET}^{\textit{LO}}$ is chosen from the interval of $[0,100]$ 
then $-\log_{10}(PFH)|_{(0<\textit{WCET}^{\textit{LO}}\leq 100)}$ $\geq$ 7, indicating higher reliability (i.e., a lower calculated PFH value) compared to the previous interval. 
As can be realized, a larger bound may require additional fault-tolerance overhead to satisfy the PFH constraint, thus limiting system utilization and real-time guarantees~\cite{koren2020fault}. Besides, a tighter bound may increase mode switching probability and reduce QoS.
Therefore, considering exposure-related reliability constraints during $\textit{WCET}^{\textit{LO}}$ selection can help reduce unnecessary reliance on costly fault-tolerance mechanisms while preserving schedulability and QoS.

To this end, rather than determining $\textit{WCET}^{\textit{LO}}$ only as a trade-off between utilization and mode switching, we argue that PFH constraints should also be incorporated into the design-time selection of execution-time bounds, i.e., $\textit{WCET}^{\textit{LO}}$. The proposed method is not intended to replace fault-tolerance techniques. Instead, it first examines whether suitable $\textit{WCET}^{\textit{LO}}$ values can satisfy conservative PFH constraints without introducing additional fault-tolerance overhead. If the PFH target cannot be met through bounding-time selection alone, the remaining gap indicates where explicit fault-tolerance mechanisms are still necessary. In this way, $\textit{WCET}^{\textit{LO}}$ selection becomes a first design layer for PFH aware MC system design, potentially reducing the amount of fault-tolerance overhead required in later design stages.

To the best of our knowledge, existing analytical schemes for determining $\textit{WCET}^{\textit{LO}}$ primarily focus on improving system timing behavior, such as processor utilization and mode-switching probability, without explicitly incorporating reliability/PFH requirements into the selection process. \textit{\proposedapp} builds on these analytical bounding-time approaches and extends them by introducing an explicit reliability-aware design layer. Under the adopted transient-fault model, each candidate $\textit{WCET}^{\textit{LO}}$ is considered as a conservative design-time exposure budget, and the resulting PFH bound is evaluated before the candidate is considered feasible. Accordingly, the PFH requirement is imposed as a hard design constraint during $\textit{WCET}^{\textit{LO}}$ determination, in addition to the real-time schedulability constraint. Therefore, a candidate $\textit{WCET}^{\textit{LO}}$ that provides a favorable trade-off between processor utilization and mode-switching probability may still be excluded if it violates the specified PFH requirement. In this way, \textit{\proposedapp} extends existing timing-oriented $\textit{WCET}^{\textit{LO}}$ selection approaches toward reliability-constrained design-time bounding.

The contributions of this paper are as follows:
\begin{itemize}
    \item Introducing PFH as an explicit design constraint in $\textit{WCET}^{\textit{LO}}$ determination for MC tasks, and show how execution-time bounds affect conservative exposure-based PFH analysis under a stated transient-fault model.
    \item Proposing \textit{\proposedapp}, a reliability-aware $\textit{WCET}^{\textit{LO}}$ determination scheme that makes a reasonable trade-off between processor utilization, mode switching probability, and QoS while satisfying real-time and PFH constraints for each criticality level.
    \item Evaluating the proposed scheme using both real and synthetic benchmarks and comparing it with state-of-the-art $\textit{WCET}^{\textit{LO}}$ selection methods, demonstrating the efficacy of PFH-aware bounding-time selection in satisfying the modeled PFH constraints while maintaining competitive utilization and mode switching probability.
    
\end{itemize}

This paper is organized as follows. Section~\ref{sec:RW} reviews related works. Sections~\ref{sec:Models} and~\ref{sec:PPMethod} present the employed models, and the proposed scheme, respectively. Section~\ref{sec:Results} reports the experimental evaluation, and Section~\ref{sec:Conclusion} concludes the paper and discusses future directions.

\section{Related Work}
\label{sec:RW}
Recently, $\textit{WCET}^{\textit{LO}}$ determination in MC systems has gained attention as a means to improve MC system timing behavior.
Early studies such as \cite{Baruah2012,Liu2016,Liu2018,Guo2018,bayati2016} simply set $\textit{WCET}^{\textit{LO}}$ as a fixed fraction of $\textit{WCET}^{\textit{HI}}$. While straightforward, this fraction-based approach often decreases utilization or leads to frequent mode switches. Despite its simplicity and ease of implementation, it lacks adaptability to task-specific execution characteristics, thereby limiting its effectiveness and frequently resulting in suboptimal system performance.

To move beyond such static approaches, researchers in \cite{Ranjbar2021Date,Ranjbar2022BotMiCS} have proposed a theoretical-based approach based on Chebyshev's theorem, to derive $\textit{WCET}^{\textit{LO}}$ to minimize the probability of mode switching at design time, thereby allowing more LC task budgets without sacrificing schedulability. \cite{ranjbar2026antimics} has proposed an analytical approach to analyze the applications and obtain $\textit{WCET}^{\textit{LO}}$ for HC tasks by defining a metric based on the application execution time distribution, in order to Enhance system utilization and QoS in the HI mode and minimize mode switches. 
Besides, researchers in~\cite{Kumar2024,Kumar2024access} applied machine-learning techniques to predict task-specific $\textit{WCET}^{\textit{LO}}$ values from code and execution data, demonstrating that careful design-time selection can significantly improve processor utilization and QoS compared to unsophisticated fraction-based assignments. \cite{Ranjbar2024} has also employed Graph-Neural-Network (GNN) technique to determine $\textit{WCET}^{\textit{LO}}$ for dependent tasks while improving QoS and reducing mode switching probability. Although these methods improve the system timing behavior, they primarily focus on utilization and mode-switching aspects. All these approaches (\cite{Ranjbar2021Date,Ranjbar2022BotMiCS,Kumar2024,Kumar2024access,Ranjbar2024,ranjbar2026antimics}) overlook reliability and do not consider how variations in $\textit{WCET}^{\textit{LO}}$ affect the PFH/reliability of tasks across different criticality levels. 
Although some works, such as \cite{Gu2016,Ranjbar2023,Ranjbar2023DATE} have explored dynamic run-time adjustment of $\textit{WCET}^{\textit{LO}}$, they also fail to explicitly account for reliability.

Moreover, \cite{Guo2022,abdeddaim2017} introduce probabilistic and adaptive techniques into MC scheduling. \cite{Guo2022} improves HC \textcolor{black}{reliability} by allocating recovery slack, but this conservative reservation reduces utilization and limits QoS. In contrast, \cite{abdeddaim2017} delays mode switches to enhance QoS, yet provides no strict PFH/reliability guarantees. Therefore, neither approach can ensure HC task reliability constraint,~even in the worst-case scenario when the system switches to HI mode.

From the perspective of QoS improvement in MC systems, most state-of-the-art works drop all/some LC tasks when the system switches to the HI mode. While this drop-based policy preserves HC task execution, it significantly reduces LC task QoS and reliability. To mitigate this strict policy, some works like~\cite{Liu2016, Liu2018} have employed degradation policy, 
which significantly improves the number of completed LC tasks compared to the dropped policy. 
However, even these approaches, though ensuring minimum QoS, often overlook how such degradation policy impacts \textcolor{black}{PFH} of HC and LC tasks.

Moreover, from the reliability perspective in MC systems, most state-of-the-art works such as~\cite{Huang2014,Ranjbar2020Access,Zeng2016,bayati2016,Choi2018}, have employed fault-tolerance techniques to improve reliability. While these approaches may meet the \textcolor{black}{PFH} constraint, 1)~they assumed a fixed $WCET^{LO}$ for HC tasks and neglect its impact on \textcolor{black}{PFH}, and 2) they incur high overheads due to applying fault-tolerance techniques.

In conclusion, prior studies have either focused on determining $\textit{WCET}^{\textit{LO}}$ to improve system timing behavior or enhancing QoS for LC tasks, typically treating reliability as a separate concern. We argue that reliability/PFH must be explicitly considered in MC system design, alongside QoS, and mode switching probability, by determining appropriate $\textit{WCET}^{\textit{LO}}$ values for HC tasks. 
This paper establishes a $\textit{WCET}^{\textit{LO}}$ determination framework that explicitly accounts for PFH/reliability constraints, ensuring that the resulting solution satisfies both real-time and PFH requirements.

\section{Models}
\label{sec:Models}
For clarity, Table~\ref{tab:notation} summarizes the main notation
used in the task, reliability, and optimization models.

\subsection{Task and System Model}

\begin{table}[t]
\centering
\caption{Main notation used throughout the paper.}
\label{tab:notation}
\begin{tabular}{c l c l}
\hline
\textbf{Symbol} & \textbf{Definition} &
\textbf{Symbol} & \textbf{Definition} \\
\hline

$\tau_i$ &
Task $i$ &
$\zeta_i$ &
Criticality level, $\zeta_i\!\in\!\{LC,HC\}$ \\

$P_i$ &
Task period &
$D_i$ &
Task deadline, $D_i=P_i$ \\

$WCET_i^{LO}$ &
LO-mode execution bound &
$WCET_i^{HI}$ &
HI-mode execution bound \\

$ACET_i$ &
Average-case execution time &
$\sigma_i$ &
Execution-time standard deviation \\

$n_i$ &
WCET selection parameter &
$P_i^{ovr}$ &
Task-overrun probability bound \\

$d_f$ &
LC degradation factor &
$U_l^m$ &
Utilization of level $l$ in mode $m$ \\

$P_{\mathrm{Sys}}^{MS}$ &
Mode-switching probability &
$\lambda$ &
Transient-fault rate \\

$Rel_i^m$ &
Reliability of $\tau_i$ in mode $m$ &
$Rel_\zeta^m$ &
Reliability of level $\zeta$ in mode $m$ \\

$Rel_\zeta$ &
Overall reliability of level $\zeta$ &
$PFH_\zeta$ &
Computed PFH bound \\

$PFH(\zeta)$ &
Required PFH constraint &
$m$ &
System mode, $m\!\in\!\{LO,HI\}$ \\

\hline
\end{tabular}
\end{table}

\label{subsec:TaskModel}

We suppose an MC system consisting of independent periodic tasks. 
Similarly to the most recent works such as~\cite{Baruah2012,Liu2016,Ranjbar2021Date}, we consider dual-criticality systems, in which the MC system supports two criticality levels of tasks, so each MC task can be classified as either HC ($\zeta$=\textit{HC}) or LC ($\zeta$=\textit{LC}). 
The specification of each task $\tau_i$ is given as a $\{\zeta_i,\textit{WCET}^{\textit{LO}}_i,\textit{WCET}^{\textit{HI}}_i,P_i,D_i\}$. 
Considering the two criticality levels of the tasks, each task is associated with two WCET values. Specifically, $\textit{WCET}_{i}^{\textit{HI}}$ and $\textit{WCET}_{i}^{\textit{LO}}$ denote the WCET assumed in the HI and LO mode, respectively. For each HC task, $\textit{WCET}_{i}^{\textit{HI}}$$\geq$$\textit{WCET}_{i}^{\textit{LO}}$~\cite{Baruah2012,Liu2016,Liu2018,Ranjbar2021Date}. 
We also consider a degradation factor $d_f$ for LC tasks, which models the reduction of the LC execution budget in HI mode rather than completely dropping LC tasks. 
The $d_f$ represents a designer specified service level for LC tasks during HI mode operation. A smaller $d_f$ corresponds to stronger service degradation and lower execution demand, while a larger $d_f$ preserves a higher level of functionality and QoS at the cost of increased processor utilization in HI mode, compared to smaller $d_f$ value~\cite{Liu2016, Liu2018}. 
We assume a uniform $d_f$ for all LC tasks to remain consistent with prior degradation-based MC models such as~\cite{Liu2016}; task-specific degradation factors are orthogonal to the proposed method and are left for future work. 
As a result, for each LC task, $\textit{WCET}_{i}^{\textit{HI}}$=$d_f\times$$\textit{WCET}_{i}^{\textit{LO}}$, where $0<d_f\leq 1$ and therefore $\textit{WCET}_{i}^{\textit{HI}}$$\leq$$\textit{WCET}_{i}^{\textit{LO}}$~\cite{Liu2016,Liu2018}. In addition, $P_i$ and $D_i$ denote the period and deadline of task, where $D_i$=$P_i$~\cite{Liu2016,Liu2018,Ranjbar2021Date}.

\textit{System Operational Model:} The MC system is initiated in the LO mode, where both HC and LC tasks are executed successfully without missing their deadlines. Whenever the execution time of at least one HC task exceeds its $\textit{WCET}^{\textit{LO}}$, and the output is not ready, the system switches to the HI mode. In this HI mode, all HC tasks execute with their $\textit{WCET}^{\textit{HI}}$, while LC tasks are degraded according to the degradation factor $d_f$. The degraded LC implementation is assumed to be available at design time and selected upon entering HI mode. Any implementation selection, data-transformation, or input-reduction overhead is assumed to be included in the corresponding HI-mode WCET budget. This ensures that HC tasks do not miss their deadlines and helps prevent potentially disastrous consequences. 
Moreover, analogous to~\cite{Liu2018,Ranjbar2022BotMiCS}, to evaluate the effect of the degradation policy on LC tasks, we use QoS metric, which is computed as the ratio of completed LC tasks to the total number of LC tasks~\cite{Ranjbar2023,Ranjbar2020Access}. 
 
In this paper, LC degradation is interpreted as reduced service quality, rather than improved functionality or improved reliability. Therefore, QoS and PFH are treated as separate quantities; QoS captures the service level delivered by degraded LC tasks, whereas PFH presents the exposure-related failure bound under the adopted transient-fault model. 
Now, the MC system switches back to the LO mode when no HC task is waiting in the processor queue to be executed~\cite{Ranjbar2021Date,Baruah2012,Liu2016,Liu2018}. The model accounts for the execution demand of degraded LC tasks through their HI mode WCETs, but does not explicitly model shared-resource interference. Such effects are assumed to be conservatively reflected in the provided WCET values and are left for future interference-aware extensions.

\subsection{Reliability Model}
There are various safety standards employed across industries, such as ISO~26262 \cite{ISO26262} for automotive applications. 
These standards establish levels of functional safety, known as Safety Integrity Levels~(SIL) in the automotive domain~\cite{Esper2015,Ernst2016}. Targeting automotive applications, ISO~26262 categorizes criticality into four levels, ASIL-A representing the lowest level and ASIL-D the highest, as detailed in TABLE~\ref{tab:SafetyStandarad}. As shown, PFH, as specified by the safety standards, is determined for each criticality level to define reliability requirements~\cite{Huang2014,Ranjbar2020Access}. This table shows the different impact on the system if the PFH constraint cannot be satisfied for tasks with different criticality levels. The criticality levels of MC systems are determined in accordance with this safety standard. 
For each criticality level~$\zeta$, we distinguish between the required PFH constraint, denoted by $PFH(\zeta)$, and the PFH bound computed under the adopted reliability model, denoted by $PFH_\zeta$.
A design is considered reliability-feasible if
$PFH_\zeta \leq PFH(\zeta)$. The reliability quantities introduced below are subsequently used in Section~\ref{subsec:ObjAnalys} to derive $PFH_\zeta$ and evaluate this constraint.

The adopted reliability model focuses on transient-fault exposure during task execution. It does not assume that changing $\textit{WCET}^{\textit{LO}}$ changes the physical fault rate or the actual execution-time distribution of a task. Instead, the mode-dependent WCET values are used as conservative design-time exposure windows to derive PFH bounds. Therefore, the PFH values computed in this paper should be interpreted as model-derived, exposure-based bounds, rather than empirically measured hardware-failure rates.

\begin{table}[t]
    \centering
    \caption{ISO 26262 safety requirement~\cite{ISO26262}.}
    \label{tab:SafetyStandarad}
    \adjustbox{max width=\linewidth}{
    \begin{tabular}{|c|cccc|}
    \hline
         $\zeta$ & ASIL-D & ASIL-C & ASIL-B & ASIL-A \\
         \hline
         \hline
         $PFH(\zeta)$ & $10^{-8}$ & $10^{-7}$ & $10^{-6}$ & $10^{-5}$ \\
         \hline
         Failure Condition & Catastrophic & Hazardous & Major & Minor\\
         \hline
    \end{tabular}
    }
\end{table}

Under the adopted constant-rate transient-fault model~\cite{koren2020fault,Koc2019,Ranjbar2026} the reliability of task $\tau_i$ in mode $l$ is defined as the
probability of fault-free execution, i.e., that no transient fault occurs during its conservative exposure interval $\textit{WCET}_i^l$. Accordingly, the task reliability is
computed as:  
\begin{equation}
    Rel_{i}^{l} = e^{-\lambda\times WCET_{i}^{l}} \quad \quad l\in\{LO,HI\}
    \label{eq:task_reliability}
\end{equation}
where $\lambda$ denotes the assumed transient-fault rate. We assume independent transient faults across task executions, consistent with commonly used exposure-based reliability analyses. Eq.~\eqref{eq:task_reliability} provides a conservative reliability bound by considering the WCET of each task in mode~$l$~as the maximum fault exposure time in that operational mode. Although the actual execution time of a task at run-time may be less than its WCET, MC systems require conservative design-time analysis under worst-case execution assumptions. Therefore, $\textit{WCET}_i^l$ is not interpreted as the actual execution time of every job, but as the design-time exposure budget used for PFH analysis~\cite{koren2020fault,Salehi2016,Zhu2004}. If exact future execution times were known, exposure could be computed from those actual executions; however, in the design-time setting, mode-dependent WCET values are used to obtain conservative PFH bounds.

Since the tasks are periodic, the one-hour PFH analysis considers all recurring activations of each task. The number of activations of task $\tau_i$ within the one-hour reference interval is determined from its period $P_i$. As tasks execute sequentially on a single-core processor and transient faults are assumed to be independent across task executions, their reliability contributions are combined multiplicatively. For compactness, Eq.~\eqref{eq:reliability_mc} is expressed at the task level and does not explicitly index individual job activations; however, each task contributes according to the number of its activations during the one-hour interval. Thus, $Rel_{\zeta}^{l}$ represents the cumulative mode-dependent reliability bound of all executions of tasks belonging to criticality level $\zeta$ over one hour~\cite{koren2020fault,Salehi2016,Zhu2004}.
\begin{equation}
    Rel_{\zeta}^{l} = \prod_{\tau_i\in\zeta}{Rel_{i}^l} \quad\zeta\in\{LC,HC\}~\&~l\in\{LO,HI\}
    \label{eq:reliability_mc}
\end{equation}

Accordingly, the selected $WCET^{LO}$ values determine the LO-mode reliability bounds through Eq.~\ref{eq:task_reliability}, while their overrun
probabilities determine the likelihood of entering HI mode. These quantities are subsequently combined in Section~\ref{subsec:ObjAnalys} to derive
the PFH bound $PFH_\zeta$ for each criticality level.
Note that the adopted PFH model focuses on execution-time exposure to transient faults and does not explicitly model fault activation, propagation, masking, recovery, or latent-error accumulation mechanisms.
The objective of this work is not to provide a complete fault-error-failure model, but rather to incorporate exposure-aware reliability constraints into design-time $\textit{WCET}^{\textit{LO}}$ determination.

\section{Proposed Method}
\label{sec:PPMethod}
In this section, first, an example is presented in Section~\ref{subsec:MotivExxp} to motivate the degraded-based scheme for LC tasks. 
Then, in Section~\ref{subsec:WCETDetermination}, we first discuss the employed scheme to determine low WCETs and the probability of overrunning them. The scheduling analysis is presented in Section~\ref{subsec:SchAnalys}. The objective analysis and optimization of the system, along with the discussion of how the \textit{\proposedapp} works, are presented in Section~\ref{subsec:ObjAnalys}.

\subsection{QoS-Aware Degradation of LC Tasks}
\label{subsec:MotivExxp}

We first present an example to motivate the use of LC task degradation in the HI mode. 
As mentioned in Section~\ref{sec:Intro}, determining the appropriate $\textit{WCET}^{\textit{LO}}$ for each HC task is a major design challenge for MC systems. The proposed scheme designs the MC systems and analyzes the MC tasks at design-time. Based on the analysis, the scheme chooses a proper $\textit{WCET}^{\textit{LO}}$ for each HC task, which improves the number of scheduled LC tasks and QoS and reduces the probability of mode switching, while guaranteeing the PFH constraint of the HC and LC tasks in both operational modes.

\begin{figure}[t]
    \centering
    \begin{minipage}[t]{1\linewidth}
        \centering
        \subfloat[Complete Version]{\includegraphics[width=0.49\columnwidth]{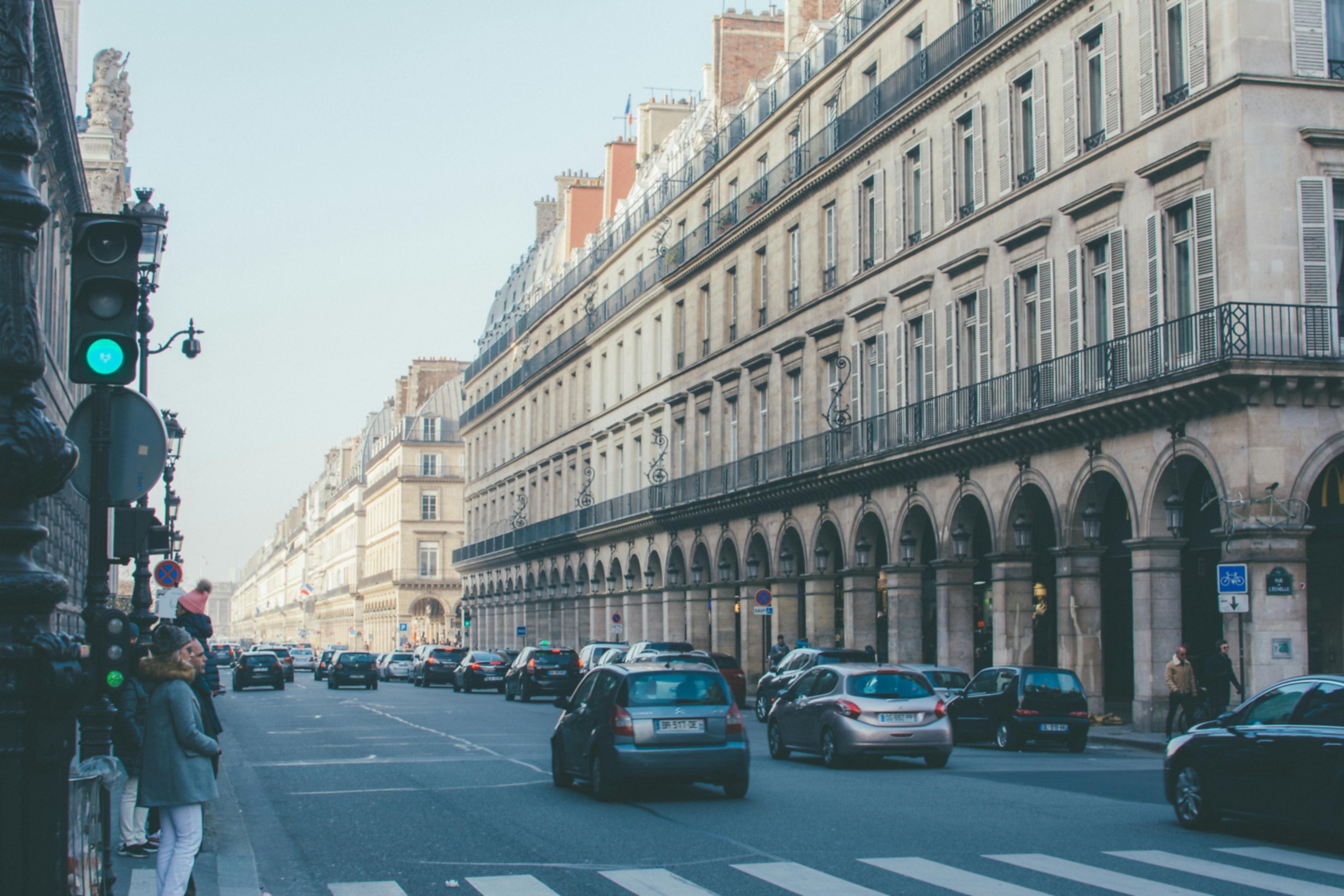}\label{fig:CV}}
        \hfill
        \subfloat[Degraded Version]{\includegraphics[width=0.49\columnwidth]{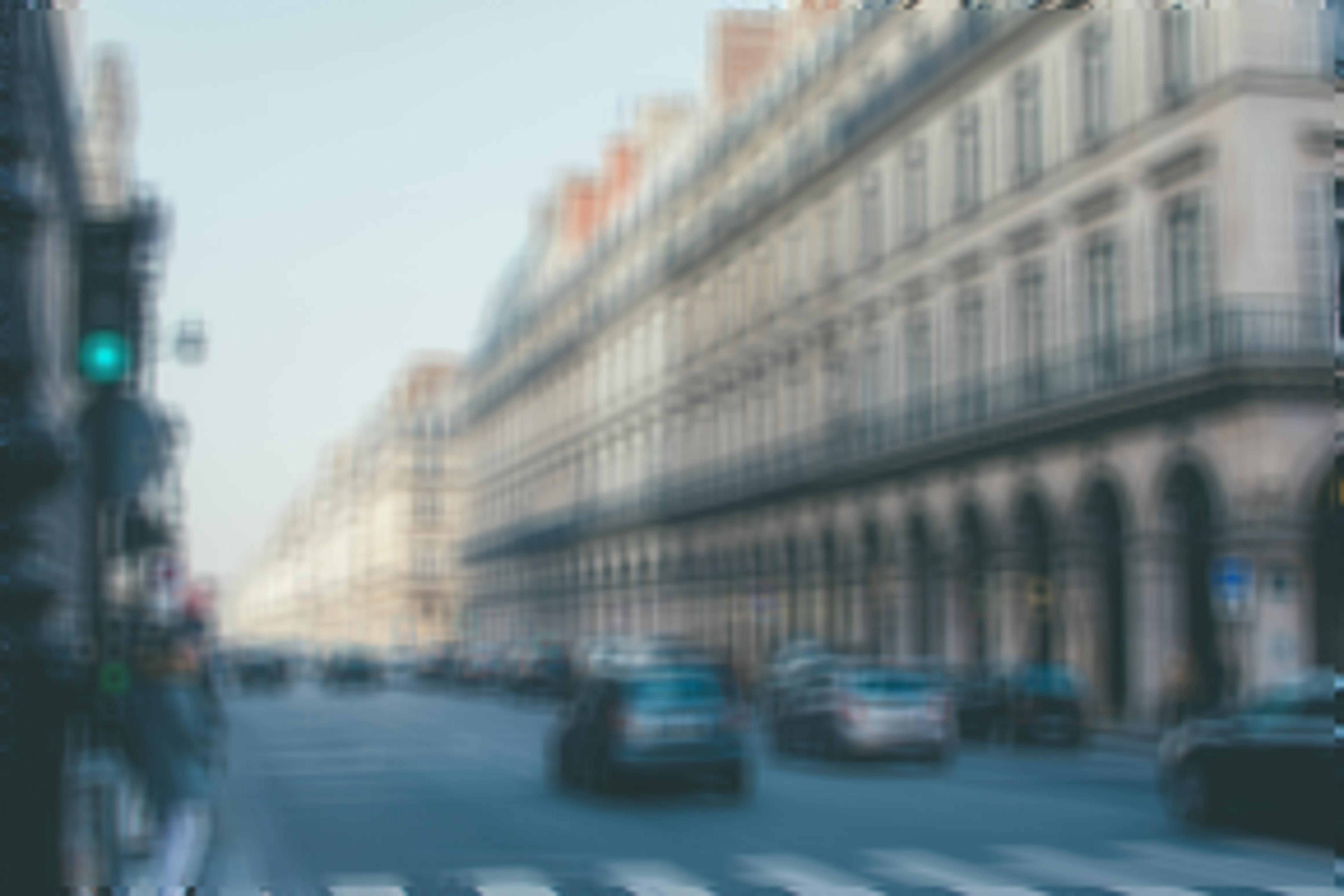}\label{fig:DV}}
        \vspace{0.8em}
        \subfloat[Policies Comparison]{\includegraphics[width=0.55\columnwidth]{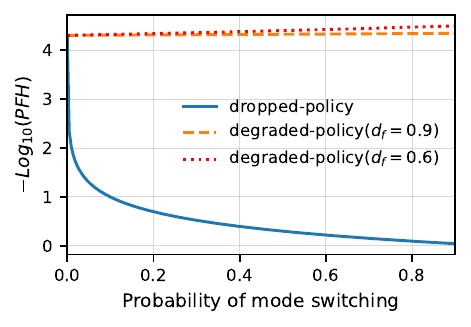}\label{fig:lc_task_reliability_pms}}
        \caption{(a)(b)~Image-processing application showing execution of the original LC task (Execution Time= 1.89~s) and its simplified version (Execution Time= 0.53~s) used in HI mode to preserve its timing reliability. (c)~PFH of LC tasks versus probability of mode switching for dropped- and degraded-policy.}
        \label{fig:lc task variants}
    \end{minipage}
\end{figure}

A key aspect of this design and making a reasonable trade-off between objectives while determining $\textit{WCET}^{\textit{LO}}$s, is the treatment of LC tasks after a mode switch. 
Considering low values of $\textit{WCET}^{\textit{LO}}$ for HC tasks may reduce the exposure budgets used in the PFH analysis and thus enhance their reliability/PFH, but increases the probability of mode switching and consequently reduces the QoS; i.e., dropping or degrading LC tasks in the HI mode. 
While completely dropping LC tasks simplifies MC task analysis, it eliminates the corresponding service and stops the LC jobs from delivering outputs before their deadlines. However, in some applications, dropping LC tasks is not acceptable because it entirely removes the associated service. 
In fact, functional failure occurs under a dropping policy. 
In contrast, degradation-based policy preserves service availability by allowing LC tasks to continue execution and still complete before their deadline, albeit with reduced functionality and execution demand, while still producing usable result.
For example, LC tasks may involve image, video, or signal processing function, which can temporarily operate at a reduced service level and produce approximate results under degraded conditions. Fig.~\ref{fig:CV} and~\ref{fig:DV} illustrate it for an LC task, using a real image-processing benchmark executed on our evaluation platform, highlighting the difference between the original algorithm and its concise variant. Although the simplified output is blurrier image than the original one, it still preserves sufficient visual information for the application. The benchmark is used in autonomous driving applications, and in conjunction with the object detection function, an HC task. Hence, the use of a degradation policy for an LC task may cause a reduction in its execution time~\cite{Liu2016,Liu2018}. In fact, lightweight variants are provided for LC tasks that are more concise than the original implementations, and to be used in HI mode.

Now, considering the degradation policy for LC tasks in HI mode, each LC task $\tau_i$ is associated with a degradation factor $d_f$, where $0<d_f\leq1$, as mentioned in Section~\ref{subsec:TaskModel}. This degradation factor represents the fraction of QoS preserved during HI mode. For example, $d_f$=0.6 indicates that 60\% of the service level is maintained. Therefore, the degraded execution budget is modeled as
$\textit{WCET}_{i}^{\textit{HI}}$=$d_f\times \textit{WCET}^{\textit{LO}}_i$. Thus, stronger degradation reduces the execution demand for LC tasks and consequently its associated processor utilization in HI mode, whereas larger values of $d_f$ present higher QoS at the expense of increased execution demand.

Fig.~\ref{fig:lc_task_reliability_pms} compares the PFH bound, derived using the adopted transient fault exposure model, for a representative task for different probability of mode switching. Three scenarios are considered: 1)~all LC tasks are dropped when the system switches to HI mode, and 2) all LC tasks are degraded rather than dropped, using two different degradation factors. 
Since degraded LC tasks execute with reduced budgets in HI mode, the resulting exposure windows are smaller than those with the original execution budgets. Consequently, the computed PFH may be lower (i.e., better). 
Hence, this observation should be interpreted as an effect of the adopted exposure-based PFH model and not as an improvement in the functional reliability of the degraded output.
As shown in Fig.~\ref{fig:lc_task_reliability_pms}, the difference between the two policies becomes more pronounced as the mode-switching probability increases. Under the dropped policy, an LC task does not execute in HI mode and therefore does not provide a successful service output, which is treated as a service failure in the adopted model. In contrast, under the degradation policy, the LC task continues to execute in HI mode with the reduced budget. Hence, increasing the likelihood of HI-mode operation gives more weight to this shorter execution budget, reducing the analyzed transient-fault exposure and consequently the computed PFH bound. 
Note that the
$d_f$=0.6 curve shows lower PFH bounds than the $d_f$=0.9 curve due to the smaller execution budget. However, this reduction in PFH comes at the cost of lower output quality, illustrating the trade-off between QoS and the exposure-based PFH metric. 
This illustrates that QoS and PFH are treated separately in this work. QoS reflects the service quality provided by LC tasks, while PFH captures the exposure-related failure bound under the adopted transient-fault model. 

\subsection{Low WCET Determination and Overrunning Probability}
\label{subsec:WCETDetermination}

To determine $\textit{WCET}^{\textit{LO}}$~for~each HC task $\tau_i$, we follow the analytical strategy used in ~\cite{Ranjbar2021Date,Ranjbar2022BotMiCS}, which is based on Chebyshev’s theorem. This formulation provides candidate $\textit{WCET}^{\textit{LO}}$ values together with an upper bound on their corresponding overrun probabilities. \textit{\proposedapp} employ this analysis to evaluate the PFH bound for each candidate, thereby restricting the subsequent optimization to candidates that satisfy both timing and reliability requirements. For each task $\tau_i$, its Average-Case Execution Time~($\textit{ACET}_i$) and standard deviation ($\sigma_i$) of the execution times are required:
\begin{equation}
    \label{eq:wcetLO}
    WCET_i^{LO} = ACET_{i} + n_i \times \sigma_{i} 
\end{equation}

Here, $n_i$ is a non-negative design-time parameter that determines how far $\textit{WCET}_i^{LO}$ is placed above $\textit{ACET}_i$. The candidate search starts from $n_i=0$ and progressively increases $n_i$ as long as the resulting $\textit{WCET}_i^{LO}$ remains smaller than $\textit{WCET}_i^{HI}$. Hence, the admissible range of $n_i$ is task-dependent and is determined by the corresponding $\textit{ACET}_i$, $\sigma_i$ and $\textit{WCET}_i^{HI}$. As a result, the probability of overrunning ($P^{ovr}_i$) for the task~$\tau_i$ is bounded by $\frac{1}{1+n_i^2}$~\cite{Ranjbar2021Date,Ranjbar2022BotMiCS}.
The parameter $n_i$ must be chosen carefully; a large value of $n_i$ increases $\textit{WCET}^{\textit{LO}}_i$, which reduces the probability of overrunning and, consequently, the probability of mode switching, 
but it increases the conservative exposure window used in the PFH analysis and may make the PFH constraint harder to satisfy. A small value of $n_i$ reduces the analyzed exposure window, but increases mode switching probability and degrades QoS. Note that the parameter $n_i$, and consequently the resulting $\textit{WCET}^{\textit{LO}}$,  is determined analytically at design-time and thus does not incur any additional run-time overhead.

\subsection{Task Schedulability Analysis}
\label{subsec:SchAnalys}
To schedule MC tasks on a single-core processor, this work uses the existing MC scheduling algorithm, Earliest Deadline First with Virtual Deadline (EDF-VD), which has become the most widely adopted MC scheduling approach in many studies~\cite{Baruah2012,Liu2016,Guo2018,Ranjbar2021Date}. 
If $U_l^m$ expresses the utilization of tasks with the same criticality level~$l$ in the mode~$m$, then $U_l^m$=$\sum_{\zeta_i\in l}{\frac{WCET^m_i}{P_i}}$, where $l$$\in$$\{LC,HC\}$, $m$$\in$$\{LO,HI\}$.

To guarantee the schedulability of MC tasks by EDF-VD, Eq.~\eqref{eq:schedulable_mc_system} must be satisfied~\cite{Ranjbar2020Access}. This equation encompasses all the necessary and sufficient conditions, needed to guarantee the MC task schedulability across all system operating modes, LO and HI, while LC tasks are degraded, and implicitly accounts for mode-switching overhead within the schedulability analysis~\cite{Baruah2012,Ranjbar2020Access,Ranjbar2023}.
\begin{equation}
    max (U_{HC}^{LO}+U_{LC}^{LO}, U_{HC}^{HI}+U_{LC}^{HI}+ \frac{U_{HC}^{LO}\times(U_{LC}^{LO} - U_{LC}^{HI})}{1-U_{LC}^{LO}}) \leq 1
    \label{eq:schedulable_mc_system}
\end{equation}

Note that although the EDF-VD scheduling algorithm is employed in this work, our scheme can be applied to any scheduling algorithm with any policy of task execution.

\subsection{Objectives Analysis and Optimization}
\label{subsec:ObjAnalys}
The system timing behavior is improved by pursuing the primary objectives discussed as follows.

\subsubsection{\textbf{Resource Utilization}}
\label{subsec:util}

Utilization is significantly improved by allowing more utilization that can be allocated to LC tasks in LO mode ($U_{LC}^{LO}$). Although maximizing $U_{LC}^{LO}$ is desirable, it is limited by schedulability constraints, which can be derived from Eq.~\eqref{eq:schedulable_mc_system}. Equation~\eqref{eq:LOmodesch} presents the condition to guarantee the task schedulability in LO mode. The upper bound of $U_{LC}^{LO}$ is determined by the parameter $n_i$ for each HC task. 
\begin{equation}
    \label{eq:LOmodesch}
    U_{LC}^{LO} \leq 1 - U_{HC}^{LO}= 1 - \sum_{\zeta_i=HC}{\left(\frac{ACET_i + n_i\times\sigma_i}{P_i}\right)}
\end{equation}

Eq.~\eqref{eq:HImodesch} shows the condition for guaranteeing the task schedulability in HI mode while there is mode switching. Here, the maximum amount of $U_{LC}^{LO}$ depends on the values of $n_i$ for each HC task and degradation factor ($d_f$). By considering the degradation policy for LC tasks in HI mode, these LC tasks execute with their reduced service version. For simplicity, we consider the execution time of all LC tasks to be equally reduced by $d_f$ (0<$d_f$<1). Note that, adaptive task-specific $d_f$ will be considered as future work. Thus, we can express $U_{LC}^{HI}$ as $U_{LC}^{HI}$= $d_{f}\times U_{LC}^{LO}$. Therefore, the utilization 
is~bounded~as:
\begin{align}
    \label{eq:HImodesch}
    &U_{HC}^{HI}+U_{LC}^{HI}+ \left(\frac{U_{HC}^{LO} \times (U_{LC}^{LO}-U_{LC}^{HI})}{1-U_{LC}^{LO}}\right) \leq 1 \nonumber \implies\\
    &U_{HC}^{HI}+d_f\times U_{LC}^{LO}+ \left(\frac{U_{HC}^{LO} \times (U_{LC}^{LO}-d_f\times U_{LC}^{LO})}{1-U_{LC}^{LO}}\right) \leq 1 \nonumber \implies\\ 
    &d_f\times (U_{LC}^{LO})^2 - (1-U_{HC}^{HI}+d_f+U_{HC}^{LO}\times(1-d_f))\times U_{LC}^{LO} - (U_{HC}^{HI}-1)\geq 0
\end{align}

By converting the last inequality in Eq.~\eqref{eq:HImodesch} into an equality and solving it, two roots (two values of $U_{LC}^{LO}$) are extracted, and we need to pick the root(s) that lie in [0,1). In addition, another $U_{LC}^{LO}$ value is obtained from Eq.~\eqref{eq:LOmodesch}. Therefore, the minimum positive $U_{LC}^{LO}$ should be selected that can be true in both inequalities.

\subsubsection{\textbf{Probability of Mode Switching}}
\label{subsec:PMS}

One of the key objectives is to minimize the probability of mode switching. Frequent switches to the HI mode leads to repeated LC tasks' degradation, which negatively impact on overall QoS~\cite{Ranjbar2021Date,Ranjbar2022BotMiCS}. 
As mentioned in Section~\ref{subsec:WCETDetermination}, $P^{ovr}_i$ denotes the probability of a task overrunning its $\textit{WCET}^{\textit{LO}}_i$. 
Since tasks are independent and executing on a single core processor, the overall probability of mode switching can be derived from Eq.~\eqref{eq:SysPMS}~\cite{Ranjbar2021Date,Ranjbar2022BotMiCS}. 
Decreasing $n_i$ reduces $\textit{WCET}_i^{\textit{LO}}$, which in turn increases $P^{ovr}_i$ and consequently increases $P_{Sys}^{MS}$. 
\begin{equation}
    \label{eq:SysPMS}
    P^{MS}_{Sys}=1-\prod_{\zeta_i\in HC}(1-P^{ovr}_{i})=1-\prod_{\zeta_i\in HC}(1-\frac{1}{1+n_i^2})
\end{equation}

\subsubsection{\textbf{System Reliability in Each Criticality Level}}
\label{subsec:RelMode}

due to existing different system operational modes, the core challenge lies in improving and ensuring the PFH/reliability of both LC and HC tasks in any operational mode.
Since the criticality levels of tasks are chosen from the safety standards of TABLE~\ref{tab:SafetyStandarad}, the PFH must be under the limit of each criticality level. For example if $\zeta=$\textit{ASIL-C}, then the $PFH_\zeta\leq PFH(\zeta)$=$10^{-7}$.
The $PFH_\zeta$ for criticality level $\zeta$ is calculated based on the reliability over one hour (Hr) system operation, as presented in Eq.~\eqref{eq:PFH_sys}~\cite{Navardi20222,koren2020fault}. 
\begin{align}
\label{eq:PFH_sys}
    &PFH_{\zeta}=(1-Rel_{\zeta})/Hr, \quad where\quad\zeta\in\{LC,HC\}  
\end{align}

The reliability for the criticality level $\zeta$ is calculated from Eq.~\eqref{eq:rel_sys}, where $Rel_{\zeta}^{LO}$ and $Rel_{\zeta}^{HI}$ are computed by Eq.~\eqref{eq:reliability_mc} and $P_{Sys}^{MS}$ by Eq.~\eqref{eq:SysPMS}. 
In the proposed analysis, each task $\tau_i$ has design-time $\textit{WCET}^{\textit{LO}}_i$ and $\textit{WCET}^{\textit{HI}}_i$ budgets, as mentioned in Section~\ref{subsec:TaskModel}. At run-time, exceeding $\textit{WCET}^{\textit{LO}}_i$ does not lead to a system failure; rather, it triggers a transition to HI mode, where HC tasks are analyzed using their $\textit{WCET}^{\textit{HI}}_i$ budgets and LC tasks are analyzed and executed with degraded execution budgets corresponding to reduced QoS level. 
Consequently, $P_{Sys}^{MS}$ is used as a design-time estimate of the likelihood that the system operates under HI-mode execution budgets in the PFH calculation. 
It should be noted that, this estimate does not imply that mode switches cause transient faults or directly affect the reliability. Instead, they only determine which execution-time budgets are used in the analysis.
\begin{equation}
\label{eq:rel_sys}
    Rel_{\zeta}= (1-P_{Sys}^{MS})\times Rel_{\zeta}^{LO} + P_{Sys}^{MS} \times Rel_{\zeta}^{HI}    
\end{equation}

Hence, when this paper refers to reducing PFH or improving related reliability, it refers to improving the computed PFH bound under the adopted MC operational model. Selecting $\textit{WCET}^{\textit{LO}}_i$ for each task $\tau_i$ does not change the physical fault rate or the actual execution-time distribution of a task; rather, it changes the LO-mode exposure budget and influences the probability with which LO or HI mode budgets are used in the PFH calculation. 
Since LO and HI modes are associated with different execution budgets and, consequently, different exposure windows, $P_{Sys}^{MS}$ is used as a design-time weighting factor to approximate the likelihood of operating under HI-mode budgets. 
Therefore, Eq.~\eqref{eq:rel_sys} is used to model a design-time PFH estimation for offline $\textit{WCET}^{\textit{LO}}$ selection and does not model HI-mode residence time or the exact number of jobs affected by each mode switch.

As can be realized, the reliability and consequently the PFH depend on the parameter $n$ of Eq.~\eqref{eq:wcetLO}, since both $P_{Sys}^{MS}$ and $Rel_{HC}^{LO}$ are computed based on this parameter. The value of $n$ directly affects $WCET^{LO}$ of HC tasks, which in turn influences $Rel_{HC}^{LO}$. 
Furthermore, as discussed, 
the $U_{LC}^{LO}$ value depends on parameter $n$. An increase in $U_{LC}^{LO}$ allows more LC tasks to be scheduled, which may negatively impact $Rel_{LC}^{LO}$ and $Rel_{LC}^{HI}$, as indicated by Eq.~\eqref{eq:reliability_mc}. 

\subsubsection{\textbf{Objective Optimization}}
\label{subsec:ObjOpt}

The proposed scheme aims to make a reasonable trade-off between $P_{Sys}^{MS}$ and $U_{LC}^{LO}$, while ensuring the PFH constraint ($PFH(\zeta)$) for each task criticality level LC and HC. To ensure the PFH constraint, it is necessary to first identify the range of acceptable $n$ values for which the PFH constraint is satisfied at each criticality level, i.e., $PFH_{\zeta}\leq PFH(\zeta)$ where $\zeta\in\{LC,HC\}$. The calculation of $PFH_{\zeta}$ has been discussed in Section~\ref{subsec:RelMode}.

\begin{figure}[t]
\centerline{\includegraphics[width=0.9\columnwidth]{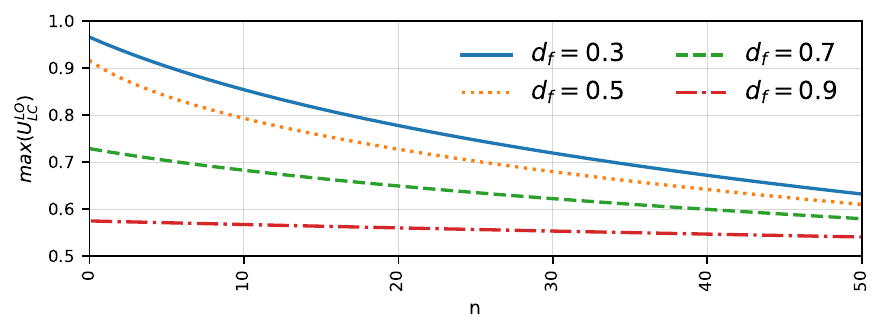}}
\caption{Variation of $U_{LC}^{LO}$ with respect to parameter $n$ and $d_f$.}
\label{fig:u_lc_lo_vs_n}
\end{figure}

Besides, as mentioned, $U_{LC}^{LO}$ varies with $d_f$ and parameter~$n$. Fig.~\ref{fig:u_lc_lo_vs_n} illustrates the impact of $n$ and $d_f$ on $U_{LC}^{LO}$. 
As parameter~$n$ increases, corresponding to a higher $\textit{WCET}^{\textit{LO}}$ for HC tasks, the utilization of HC tasks increases, thereby reducing the utilization assigned to LC tasks.
In addition, as the degradation factor increases, LC tasks are executed in HI mode with higher QoS. However, this restricts task schedulability in HI mode according to Eq.~\eqref{eq:HImodesch}, thereby reducing the impact of parameter $n$ on $U_{LC}^{LO}$. 
To achieve higher utilization, smaller values of $n$ with lower degradation factors are preferred. Since the minimum QoS (i.e., degradation factor) is specified by the designer, we should introduce an optimization function that considers the degradation factor. 
To this end, given the objectives, constraints and parameters, the optimization function is defined as Eq.~\eqref{eq:ObjOpt}, where an increase in $d_f$, reduces the weight of $U_{LC}^{LO}$ while increasing the weight of (1-$P_{Sys}^{MS}$). 
\begin{align}
\label{eq:ObjOpt}
    &maximize (\frac{1-d_f}{2} \times U_{LC}^{LO} + \frac{1+d_f}{2} \times (1-P_{Sys}^{MS})), \nonumber \\
    &PFH_{\zeta}\leq PFH(\zeta) \quad \zeta\in\{LC,HC\} 
\end{align}

We have to find the optimal value of parameter $n$ (i.e., the proper $WCET^{LO}$ for each HC task) within the selected range that ensures that PFH is met, and subsequently maximizes the optimization function, by minimizing $P_{Sys}^{MS}$ and maximizing $U_{LC}^{LO}$, while ensuring the PFH constraints. 
Note that a larger $d_f$ diminishes the influence of $n$ on $U_{LC}^{LO}$, thereby reducing the significance of optimizing $U_{LC}^{LO}$ through the selection of an optimal $n$.
Now, in order to determine the optimum $n$ over the specified range, a bounded brute-force search is employed. For each candidate value of $n$, the schedulability conditions, PFH/reliability, and the optimization function are evaluated across all tasks, resulting in a computational complexity of $\operatorname{O}(K T)$, where $T$ denotes the number of tasks and $K$ denotes the number of candidate values of $n$. 
This procedure is performed entirely offline during system design, and the resulting parameters are fixed at run-time. Therefore, the brute-force search imposes no run-time overhead.
Since $K$ is typically limited, the procedure scales linearly with the number of tasks. As a result, given the small candidate set and offline execution, the approach is both tractable and appropriate for the considered task sets. If no candidate value of $n$ satisfies both the schedulability and the PFH constraints, the task set is declared infeasible under the selected fault rate assumption, $d_f$, and PFH target. In such a case, the designer must either reduce LC load, relax the required QoS level, adjust the PFH target, or introduce explicit fault-tolerance mechanisms.

These relationships collectively highlight how critical design decisions such as selecting an appropriate $WCET^{LO}$ for HC tasks, and adjusting $d_f$ to tune the QoS, can significantly enhance the system timing behavior and reliability.
By carefully tuning these parameters, designers can achieve an effective balance between functionality, and performance, ensuring that LC tasks maintain acceptable service levels under varying operational modes, while HC tasks consistently meet stringent safety requirements.

\section{Experimental Results}
\label{sec:Results}
This section evaluates \textit{\proposedapp} in terms of utilization, mode-switching probability, QoS, and PFH using both real embedded benchmarks and synthetic task sets.

\subsection{Evaluation with Real Benchmark Model}
We conducted experiments using various real benchmarks from the MiBench suite~\cite{Mibench} including automotive and networking, and the AXBench suite~\cite{axbench}. In this subsection, we have considered \guilsinglleft insert-sort\guilsinglright, \guilsinglleft qsort\guilsinglright, \guilsinglleft bitcount\guilsinglright, \guilsinglleft edge\guilsinglright, \guilsinglleft epic\guilsinglright~and \guilsinglleft FFT\guilsinglright~as HC tasks. To obtain the WCET, ACET, and standard deviation of each benchmark, we adopt the methodology and values reported in~\cite{Ranjbar2022BotMiCS}. 
In fact, the experimental platform, ODROID XU4, is used to find and characterize task timing, including execution-time distributions to obtain timing parameters.

We compare the results of our scheme with existing MC strategies for determining $\textit{WCET}^{\textit{LO}}$, such as~\cite{Liu2016,Ranjbar2021Date} considering LC tasks are degraded in the HI mode. 
As mentioned in Section~\ref{sec:RW}, since most MC research works~\cite{Liu2016,Baruah2012,Liu2018,Guo2018,bayati2016} have considered the same policy (i.e., considering a fraction of $\textit{WCET}^{\textit{HI}}$ as $\textit{WCET}^{\textit{LO}}$), we choose~\cite{Liu2016} as a representative approach to carry out the experiments, in which for $\alpha=\frac{\textit{WCET}^{\textit{LO}}}{\textit{WCET}^{\textit{HI}}}$, we have evaluated with $\alpha\in[0.5,1]$ and $\alpha\in[0.1,1]$. 

Besides, we have considered the fault rate $\lambda=10^{-12}faults/\mu s$ based on the real fault rate considered in many studies such as~\cite{Zhu2004,Salehi2016,Huang2014}. 
The reported PFH values are not measured hardware-failure rates from the real platform. Instead, they are analytically derived from the adopted transient-fault model using the measured execution-time characteristics and assumed transient-fault parameters. 
We have considered the HC tasks from the ASIL-C level in TABLE~\ref{tab:SafetyStandarad} to determine the PFH constraint (i.e., $PFH_{HC}$$\leq$$10^{-7}$) to show the efficacy of the proposed method. 
Note that although $PFH(\zeta)$=$10^{-7}$ is used in the experiments, the proposed approach is general and employed for any PFH requirements specified by the target application.
In addition, researchers in~\cite{Ranjbar2021Date} dropped all LC tasks when the system switched to HI mode. Therefore, to ensure a fair comparison in determining the appropriate $WCET^{LO}$ policy, we consider that in~\cite{Ranjbar2021Date}, LC tasks are degraded in HI mode. Moreover, we have considered the same degradation factor of LC tasks in HI mode, for all approaches 
for fair comparison.

\begin{table}[t]
\centering
\caption{Comparison of system performance for real applications across different approaches under a PFH constraint of $10^{-7}$ ($PFH(HC)\leq 10^{-7}$).}
\label{tab:results}
\adjustbox{max width=\linewidth}{
\begin{tabular}{l | c c c c c c}
\toprule
Method & $d_f$ & $\quad P_{SYS}^{MS}$ (\%)$\quad$ & $max(U_{LC}^{LO})$$\quad$ & \textit{Opt. Obj.}$\quad$ & calculated $PFH_{HC}$ \\
\midrule
\textit{\proposedapp}
 & 0.3 & 3.3 &  0.772 &  0.899& $10^{-7}$  \\
 & 0.9 & 1.4 & 0.486 & 0.961 &  $10^{-7}$ \\
\midrule
\cite{Ranjbar2021Date}
& 0.3 &  6.2 &  0.807& 0.892 &  $10^{-7}$ \\
 & 0.9 & 1.4 & 0.485 & 0.961 &  $10^{-6}$ \\
\midrule
\cite{Liu2016}- $\alpha\in[0.1,1] $ 
& 0.3 & 4.8 & 0.576 &  0.820 &   $10^{-5}$ \\
 & 0.9 & 4.8 & 0.471 & 0.928 &  $10^{-5}$ \\
\midrule
\cite{Liu2016}- $\alpha\in[0.5,1] $ 
& 0.3 & 0.2 & 0.475 & 0.814 &  $10^{-5}$ \\
 & 0.9 & 0.2 & 0.455  & 0.970 &  $10^{-5}$ \\
\bottomrule
\end{tabular}
}
\end{table}

TABLE~\ref{tab:results} presents the results 
of mentioned benchmarks for different approaches. Here, we consider $d_f$=0.3 and $d_f$=0.9 to evaluate two different degradation scenarios: strong degradation with low execution demand, and mild degradation with higher QoS preservation. These values are not tied to a specific application requirement. In fact, they are chosen to show the sensitivity of the proposed method to different service levels of LC tasks. In practice, $d_f$ should be determined based on application-specific QoS requirements.
As illustrated, although \cite{Liu2016} and \cite{Ranjbar2021Date} can improve the \textcolor{black}{PFH} of LC tasks through degradation in HI mode, they are unable to guarantee the PFH constraints of HC tasks.
In contrast, our scheme can make a reasonable trade-off between $P_{Sys}^{MS}$ and $U_{LC}^{LO}$, while ensuring $PFH_\zeta$ being lower than the constraint ($PFH(HC)$=$10^{-7}$), compared to other approaches.
The differences in $PFH_{HC}$ across methods arise because each approach selects different $\textit{WCET}^{\textit{LO}}$ values. These values change the LO-mode execution-time budget and the probability of mode switching, which are then used in the PFH calculation. Thus, Table 2 compares the PFH feasibility of different $\textit{WCET}^{\textit{LO}}$ -selection policies under the proposed design-time model, rather than measured differences in the physical fault behavior of the platform.

As shown, for $d_f$=0.3, although \textit{\proposedapp} reduces $U_{LC}^{LO}$ by 4.3\% compared to \cite{Ranjbar2021Date}, it achieves a lower $P_{Sys}^{MS}$ (2.9\%). The optimized objective (\textit{Opt. Obj.}, Eq.~\eqref{eq:ObjOpt}) thus demonstrates a better trade-off. The advantage of \textit{\proposedapp} becomes more evident at higher QoS levels. For $d_f$=0.9, although both \textit{\proposedapp} and \cite{Ranjbar2021Date} yield nearly the same $P_{Sys}^{MS}$ and $U_{LC}^{LO}$, only \textit{\proposedapp} could satisfy the PFH constraint by explicitly accounting for PFH in determining $\textit{WCET}^{\textit{LO}}$ for HC tasks. 
Moreover, in \cite{Liu2016}, varying $d_f$ has no effect on $P_{Sys}^{MS}$ due to fixed $\textit{WCET}^{\textit{LO}}$ for HC tasks, and only a slight impact on $U_{LC}^{LO}$. This further confirms that the fraction-based $\textit{WCET}^{\textit{LO}}$ selection limits the ability to adapt to different QoS requirements. 
Although \textit{\proposedapp} has a higher $P_{Sys}^{MS}$ than \cite{Liu2016} with $\alpha$$\in$$[0.5,1]$, it achieves a significant improvement in $U_{LC}^{LO}$, resulting in a better trade-off between objectives as reflected in the \textit{Opt. Obj.} column. Notably, for any value of $d_f$, \cite{Liu2016} fails to meet the PFH constraint. Overall, the real-benchmark results show that \textit{\proposedapp} preserves
a competitive utilization/mode-switching trade-off while, unlike the compared $\textit{WCET}^{\textit{LO}}$-selection approaches, explicitly maintaining the specified PFH constraint.

\subsection{Evaluation with Synthetic Task Sets}

We conduct an extensive evaluation using synthetic task sets to assess the efficacy of \textit{\proposedapp}. 
To this end, we used a Python script to generate 1000 random MC task sets for each utilization bound consistent with prior works~\cite{Ranjbar2021Date,Ranjbar2022BotMiCS}, over the range [0.05,1] with increments of 0.05. As in the previous sub-section, we set $\lambda$=$10^{-12}faults/\mu s$. 
For each HC task, $\textit{WCET}^{\textit{HI}}$, ACET, and $\sigma$ in varying proportions are randomly generated from the ranges reported in~\cite{Ranjbar2022BotMiCS}, which are derived from real application model, given embedded applications, where tasks and their characteristics are typically known at design time~\cite{Ranjbar2022BotMiCS}.

\subsubsection{\textbf{Analyzing the Efficacy of \textit{\proposedapp}}:}

\begin{figure}[t]
\centering
\subfloat[Varying WCET Ranges]{\includegraphics[width=0.49\linewidth]{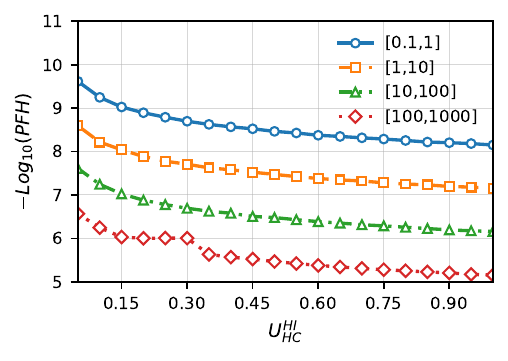}\label{fig:WCETVAary}}
\hfill
\subfloat[Varying Task Utilization]{\includegraphics[width=0.49\linewidth]{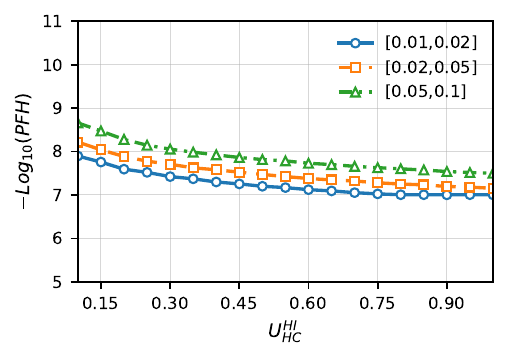}\label{fig:UtilVary}}
\caption{Sensitivity of the HC-task reliability level $-\log_{10}(PFH)$ to (a) different $WCET^{HI}$ ranges (ms) and (b) different per-task utilization ranges.}
\label{fig:reliabilityanalysis}
\end{figure}

\begin{figure}[t!]
\centering
\subfloat[$d_f$=0.3]{\includegraphics[width=0.49\linewidth]{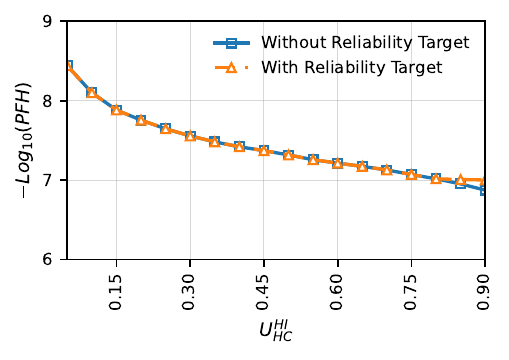}\label{fig:DegFactor1}}
\vspace{0.5em}
\subfloat[$d_f$=0.5]{\includegraphics[width=0.49\linewidth]{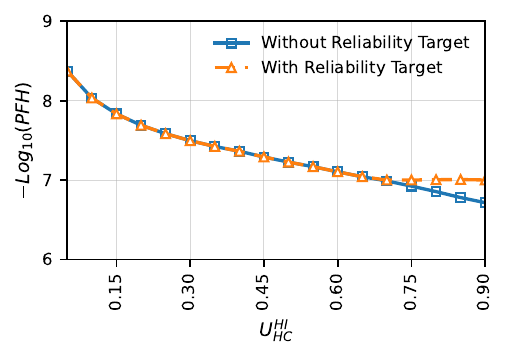}\label{fig:DegFactor2}}
\hfill
\subfloat[$d_f$=0.7]{\includegraphics[width=0.49\linewidth]{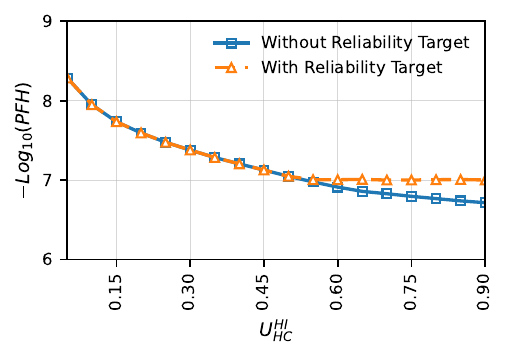}\label{fig:DegFactor3}}
\hfill
\subfloat[$d_f$=0.9]{\includegraphics[width=0.49\linewidth]{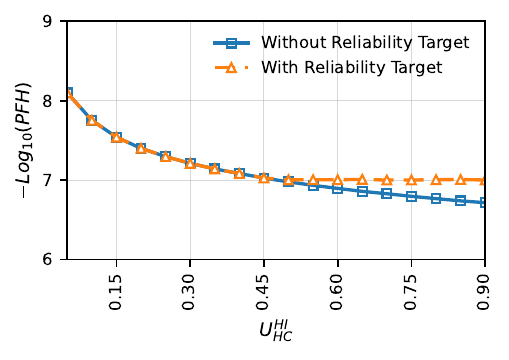}\label{fig:DegFactor4}}
\caption{HC-task reliability level $-\log_{10}(PFH)$ for different degradation factors~$d_f$, comparing $WCET^{LO}$ selection with and without enforcing the PFH constraint.}
\label{fig:degfactorVary}
\end{figure}

\begin{itemize}
    \item \textit{Analyzing the impact of $\textit{WCET}^{\textit{HI}}$ on PFH of HC tasks}: From Eq.~\eqref{eq:task_reliability}, increasing $\textit{WCET}^{\textit{HI}}$ leads to an exponential decrease in the reliability of each HC task, thereby reducing $Rel_{HC}$. Figure~\ref{fig:WCETVAary} shows the impact of varying $\textit{WCET}^{\textit{HI}}$ (in units of \textit{ms}) on PFH. In this evaluation, we consider four different ranges of $\textit{WCET}^{\textit{HI}}$ for HC tasks. As expected, higher ranges of $\textit{WCET}^{\textit{HI}}$ lead to an increase in PFH consistent with Eq.~\eqref{eq:PFH_sys}. In practice, task execution times cannot be restricted to a single range and may instead span multiple~ranges. Since these ranges correspond to different $PFH_\zeta$, such mixing highlights the strength of \textit{\proposedapp} in better meeting PFH constraint ($PFH(\zeta)$), even when the target constraint lies between $PFH_\zeta$ values.
    \item \textit{Analyzing the impact of task utilization on PFH of~HC~tasks}: 
    Figure~\ref{fig:UtilVary} shows how different task utilization ranges influence PFH. As each task utilization increases, fewer tasks can be feasibly scheduled. According to Eq.~\eqref{eq:reliability_mc}, it leads~to~an increase in reliability, thereby improving the overall reliability/PFH, according to Eq.~\eqref{eq:rel_sys}. In addition,~for constraint $PFH(\zeta)$=$10^{-7}$, the smaller task utilization range ($u_i$$\in$[0.01,0.02]) results in PFH violations without our approach for $U_{HC}^{HI}$$\geq$$0.7$. However, \textit{\proposedapp} can successfully guarantee compliance with the constraint $PFH(\zeta)$.
    \item \textit{Analyzing the impact of varying $d_f$}: 
    Fig.~\ref{fig:degfactorVary} illustrates the~effect of varying $d_f$ on PFH under two scenarios: with and without considering PFH constraint while determining  $\textit{WCET}^{\textit{LO}}$. As shown, when PFH constraint is not considered in determining $\textit{WCET}^{\textit{LO}}$ for HC tasks, the PFH constraint ($-Log_{10}(PFH)$=7) is violated.
    For smaller $d_f$, (e.g.,~Fig.~\ref{fig:DegFactor1}), the violations occurs at $U_{HC}^{HI}$=0.8, whereas for larger $d_f$ (e.g.,~Fig.~\ref{fig:DegFactor4}) the violation occurs earlier, at $U_{HC}^{HI}$=0.5. In contrast, when the PFH constraint is explicitly considered in determining $\textit{WCET}^{\textit{LO}}$, \textit{\proposedapp} never violates the constraint. Furthermore, the variation in $PFH_\zeta$ with different $d_f$ is due to the direct impact of $d_f$ on the optimization function (Eq.~\eqref{eq:ObjOpt}), which in turn affects the determination of $\textit{WCET}^{\textit{LO}}$ for HC tasks, and consequently their PFH ($PFH_\zeta$).
\end{itemize}

\begin{figure*}[t]
\centering
\subfloat[$P_{Sys}^{MS}$ (\%)]{\includegraphics[width=0.33\linewidth]{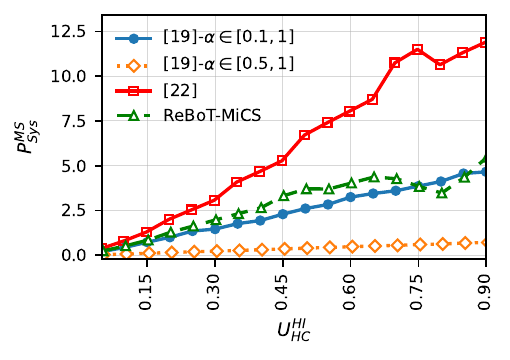}\label{fig:PMS0.3}}
\vspace{0.7em}
\subfloat[Utilization ($max(U_{LC}^{LO})$)]{\includegraphics[width=0.33\linewidth]{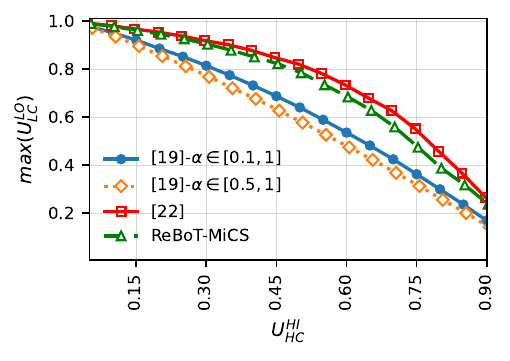}\label{fig:Util0.3}}
\hfill
\subfloat[Optimized Objective]{\includegraphics[width=0.33\linewidth]{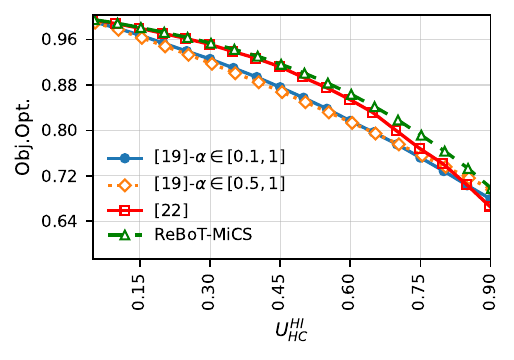}\label{fig:Obj0.3}}
\hfill
\subfloat[PFH for HC tasks]{\includegraphics[width=0.32\linewidth]{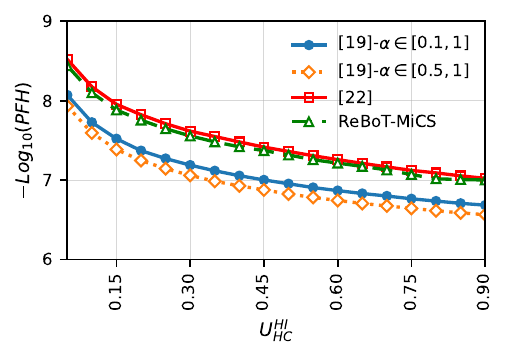}\label{fig:HCRel0.3}}
\hfil
\subfloat[PFH for LC tasks]{\includegraphics[width=0.34\linewidth]{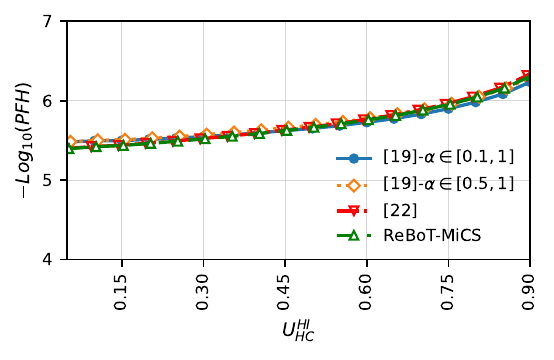}\label{fig:LCRel0.3}}
\caption{Comparing with other policies by considering $d_f=0.3$.}
\label{fig:comparing0.3}
\end{figure*}

\subsubsection{\textbf{Comparison of \textit{\proposedapp} with Other Policies}:}

For comparison, we also generate 1000 random MC task sets for each utilization bound.
Building upon the conclusions of the previous subsection, for each HC task, $\textit{WCET}^{\textit{HI}}$ is randomly generated from the interval [5, 20]\textit{ms}, covering two $\textit{WCET}^{\textit{HI}}$ ranges from Fig.~\ref{fig:WCETVAary}. The corresponding ACET, and standard deviation are randomly generated from [100, 500] and [100, 200] $\mu$\textit{s}, respectively. Each task utilization is randomly generated within the range [0.02, 0.05]. The period of each task $\tau_i$ is then computed from $\textit{WCET}^{\textit{HI}}$ and task utilization to ensure consistency with the target system utilization.

Fig.~\ref{fig:comparing0.3} 
illustrates the comparison of \textit{\proposedapp} against \cite{Ranjbar2021Date,Liu2016} in terms of objectives when $d_f$=0.3. 
Compared to~\cite{Ranjbar2021Date}, although $U_{LC}^{HI}$ is slightly lower in \textit{\proposedapp}, $P_{Sys}^{MS}$ is improved by up to 7.67\%, as reflected in the optimization function value shown in Fig.~\ref{fig:Obj0.3}. From the previous results, since $d_f$ is low, both \cite{Ranjbar2021Date} and \textit{\proposedapp} could meet the PFH constraint. Besides, due to the degradation policy used in~\cite{Ranjbar2021Date,Liu2016} for fair comparison, the PFH of LC tasks remains nearly the same across all methods. 
Furthermore, in \cite{Liu2016}, using a smaller lower-bound value of $\alpha$ (e.g., $[0.1,1]$ compared to $[0.5,1]$) increases $max(U_{LC}^{LO})$, but at the cost of higher $P_{Sys}^{MS}$. However, the PFH in \cite{Liu2016} with $\alpha$$\in$$[0.1,1]$ ($[0.5,1]$) is violated when $U_{HC}^{HI}$$>$0.45 (0.35), even though it achieves a better $P_{Sys}^{MS}$ compared to \textit{\proposedapp}. The reason for having a worse PFH value in \cite{Liu2016} with $\alpha$$\in$$[0.5,1]$, compared to $\alpha$$\in$$[0.1,1]$  arises from the higher lower-bound value of $\textit{WCET}^{\textit{LO}}$. This highlights that increasing $\textit{WCET}^{\textit{LO}}$ to reduce mode switching inherently degrades reliability, as formalized in Eq.~\eqref{eq:task_reliability}, thereby revealing a structural trade-off that fixed-fraction approaches are unable to resolve.
To this end, when $d_f$=0.3, \textit{\proposedapp} improves $max(U_{LC}^{LO})$ by up to 27.6\% (11.67\% on average), while satisfying the PFH constraints, and maintaining the $P_{Sys}^{MS}$ by~up~to~5.48\%. Furthermore, this demonstrates that incorporating PFH into the selection of $\textit{WCET}^{\textit{LO}}$ selection enables a more balanced allocation of system utilization and mode switching probability, while preserving both reliability and real-time constraints. 

\begin{figure*}[t]
\centering
\subfloat[$P_{Sys}^{MS}$ (\%)]{\includegraphics[width=0.33\linewidth]{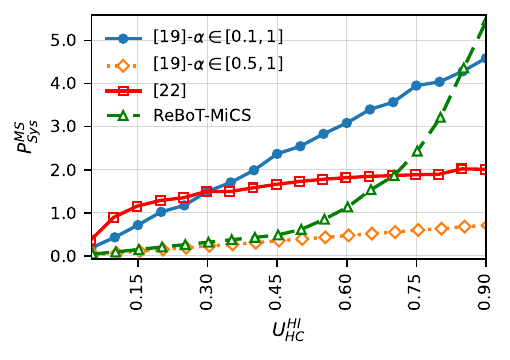}\label{fig:PMS0.9}}
\vspace{0.7em}
\subfloat[Utilization ($max(U_{LC}^{LO})$)]{\includegraphics[width=0.33\linewidth]{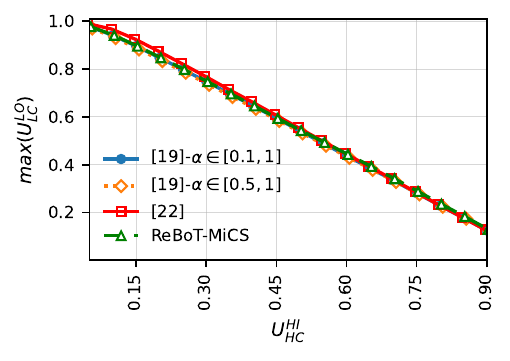}\label{fig:Util0.9}}
\hfill
\subfloat[Optimized Objective]{\includegraphics[width=0.33\linewidth]{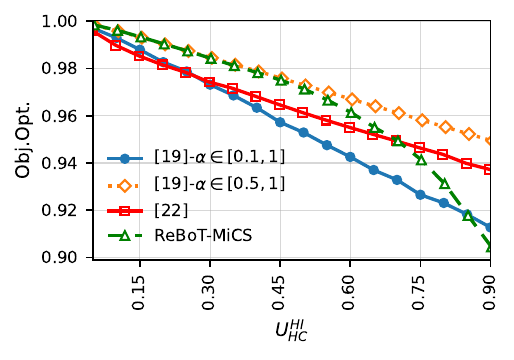}\label{fig:Obj0.9}}
\hfill
\subfloat[PFH for HC tasks]{\includegraphics[width=0.32\linewidth]{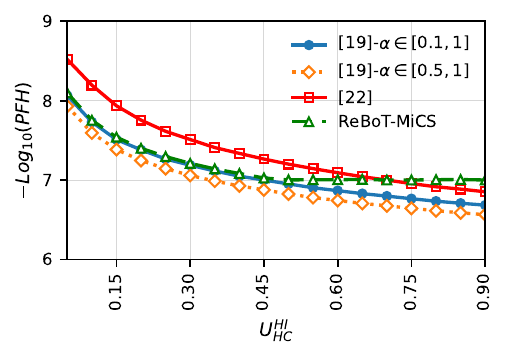}\label{fig:HCRel0.9}}
\hfil
\subfloat[PFH for LC tasks]{\includegraphics[width=0.34\linewidth]{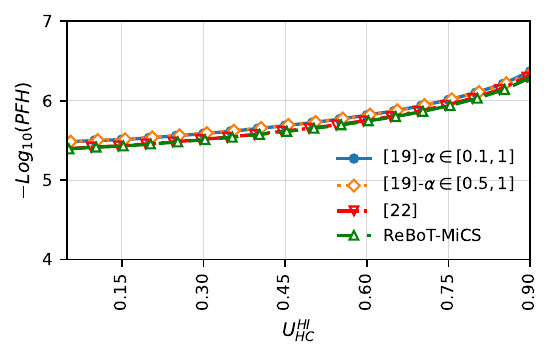}\label{fig:LCRel0.9}}
\caption{Comparing with other policies by considering $d_f=0.9$.}
\label{fig:comparing0.9}
\end{figure*}

Fig.~\ref{fig:comparing0.9} presents the same result for $d_f$=0.9. As shown in Fig.~\ref{fig:PMS0.9}, \textit{\proposedapp} sometimes yields a higher $P_{Sys}^{MS}$ than other approaches, for example, at $U_{HC}^{HI}$>$0.7$ compared to \cite{Ranjbar2021Date}.
This observation is important because, beyond this threshold, \cite{Ranjbar2021Date} violates the PFH constraint (Fig.~\ref{fig:HCRel0.9}). 
Although \textit{\proposedapp} produces a lower PFH value compared to \cite{Ranjbar2021Date} for $U_{HC}^{HI}$<$0.7$, this stems from its lower $P^{MS}_{Sys}$ and it still satisfies the PFH constraint. This suggests that a controlled increase in $P_{Sys}^{MS}$ can be acceptable when the selected $\textit{WCET}^{\textit{LO}}$ values still satisfy the PFH constraint while preserving schedulability and QoS.
Moreover, as shown in Fig.~\ref{fig:Util0.9}, all methods achieve nearly the same $max(U_{LC}^{LO})$, which results from the higher $d_f$~(i.e., QoS), limiting the upper-bound of $max(U_{LC}^{LO})$ according to Eq.~\eqref{eq:ObjOpt}. Note that the lower optimization function value of \textit{\proposedapp} compared to other approaches for certain $U_{HC}^{HI}$ is due to the inclusion of the PFH constraint in Eq.~\eqref{eq:ObjOpt}. In this case, the feasible solution space is constrained by PFH requirements, which reduces optimization flexibility but ensures reliable system operation.
To this end, considering high QoS, \textit{\proposedapp} can always satisfy the PFH constraints, and maintain the $P_{Sys}^{MS}$ by up to 5.48\% 
across the entire range of $U_{HC}^{HI}$, whereas the other approaches do not.
Overall, the synthetic-task-set results show that explicitly constraining $\textit{WCET}^{\textit{LO}}$ selection by PFH prevents reliability-infeasible solutions across the evaluated utilization and degradation ranges, while maintaining a competitive trade-off between LC utilization and mode-switching probability.

\section{Conclusion and Future Work}
\label{sec:Conclusion}
This paper introduced a reliability-aware low WCET determination scheme for MC systems that improves system timing behavior at design-time by balancing utilization that can be assigned to LC tasks, mode switching probability, and QoS, while ensuring real-time and reliability constraints. The evaluation with real and synthetic benchmarks demonstrated the efficacy of the scheme in ensuring reliability, enhancing utilization, and maintaining mode switching probability in worst-case scenario. In addition, it showed the importance of reliability-aware low WCET determination in the design of reliable MC systems.

As future work, we would extend the framework to support task-specific degradation factor rather than relying on a uniform one. Furthermore, we would enhance the work through the use of ML to determine adaptive low WCET based on run-time behavior changes. Additionally, we would integrate the scheme with fault-tolerance techniques to further strengthen system reliability while reducing redundancies. Furthermore, this work focuses on a single-core processor. While this assumption is common in MC research area and enable controlled evaluation of PFH-aware $WCET^{LO}$ selection, extension of framework to support multi-core heterogeneous platforms remains an important direction for future work, in which the WCETs of tasks may differ across cores and additional opportunities for incorporating fault-tolerance techniques arise.

\subsubsection{\ackname} This work was supported in part by Deutsche Forschungsgemeinschaft~(DFG) through the Project \textit{LeanMICS} under Grant 534919862.

\clearpage

\bibliographystyle{splncs04}
\bibliography{Ref.bib}
\end{document}